\documentclass[aps,pra,twocolumn,groupedaddress,superscriptaddress]{revtex4}
 \usepackage{lipsum}
\usepackage{mathtools,amssymb,lipsum}
\usepackage{graphicx}
\usepackage{xcolor,ulem}

\newcommand{\be}{\begin{equation}}
\newcommand{\ee}{\end{equation}}
\newcommand{\beq}{\begin{eqnarray}}
\newcommand{\eeq}{\end{eqnarray}}
\newcommand{\bea}{\begin{array}}
\newcommand{\eea}{\end{array}}

\draft 
\begin{document}

\title{Probing the optical chiral response of single nanoparticles with optical tweezers}
\author{R. Ali}
\email[]{r.ali@if.ufrj.br}
\affiliation{%
Instituto de F\'isica, Universidade Federal do Rio de Janeiro, Caixa Postal 68528, Rio de Janeiro, RJ, 21941-972, Brasil}

\author{ R. S. Dutra}
\affiliation{LISComp-IFRJ, Instituto Federal de Educa\c{c}\~ao, Ci\^encia e Tecnologia, Rua Sebasti\~ao de Lacerda, Paracambi, RJ,26600-000, Brasil}
\author{F. A. Pinheiro}
\affiliation{%
Instituto de F\'isica, Universidade Federal do Rio de Janeiro, Caixa Postal 68528, Rio de Janeiro, RJ, 21941-972, Brasil}
\author{ F. S. S. Rosa}
\affiliation{%
Instituto de F\'isica, Universidade Federal do Rio de Janeiro, Caixa Postal 68528, Rio de Janeiro, RJ, 21941-972, Brasil}
\author{P. A. Maia Neto}
\affiliation{%
Instituto de  F\'isica, Universidade Federal do Rio de Janeiro, Caixa Postal 68528, Rio de Janeiro, RJ, 21941-972, Brasil}

\date{\today}

\begin{abstract}
We propose an enantioselective scheme to sort homogeneous chiral particles using optical tweezers. For a certain range of material parameters, we show that a highly focused circularly-polarized laser beam traps particles of a specific chirality selected by the handedness of the trapping beam.  Furthermore, by applying a transverse Stokes drag force that displaces the trapped particle off-axis, we allow for the rotation of the particle center-of-mass around the trapping beam axis. 
The rotation angle is highly dependent on the handedness of the trapped particle and is easily measurable with standard video-microscopy techniques, allowing for an alternative mechanism for chiral resolution. Our platform not only allows for enantioselection of particles dispersed in solution but also paves the way to the characterization of the  chiral parameter of individual, homogeneous chiral microspheres using optical tweezing. 
\end{abstract}
\maketitle 

\section{Introduction}
Chiral structures are ubiquitous in nature and play a crucial role in the biological functions of living species. Natural chiral substances have also fundamental technological and industrial importance as they are present in many chemical and pharmaceutical compounds~\cite{wagniere}. Natural optical activity is the hallmark optical effect exhibited by chiral media, as discovered in the pioneering work by Arago, and it is typically weak for naturally occurring materials~\cite{wagniere}. With advent of metamaterials and plasmonics, new artificial media with much higher chiral optical response have been developed to generate novel optical properties and applications such as negative index media~\cite{pendry2004,plum}, broadband circular polarizers~\cite{gansel2009}, and enantioselective photochemistry~\cite{tang2011}. An important class of artificial chiral media is composed of plasmonic nanoparticles, which includes chiral nanocrystals, DNA-assembled plasmonic nanostructures, chiral plasmonic particles assembled on scaffolds, core-shell plasmonic spheres~\cite{cpreview1, cpreview2, cpreview3}, and assemblies of nanoparticles in chiral configurations, such as helices~\cite{fan2010} or even random arrangements~\cite{pinheiro2017}. {However, it is quite challenging to probe individual chiral optical response for single chiral plasmonic nanoparticles}, as the existing enantioselective methods usually can only measure the average chiral optical response of a large number of (typically) non-identical particles in solution (e.g. circular dichroism). By the same token, no established method exists to determine the chiral parameter $\kappa$ of individual nanoparticles, which would be of great interest to characterise optical chiral response in chiral nanoplasmonics~\cite{dionnereply,Mastroianni}.

To circumvent these limitations, we have recently put forward an alternative chiral resolution method to not only achieve enantioselection of chiral nanoparticles but also to determine their chiral parameter $\kappa$~\cite{rali2020}. This method is based on optical tweezing of dielectric spheres coated with chiral shells, where it is possible to select the handedness of the trapped coated particles by choosing the appropriate circular polarization of the trapping laser beam. 
Our approach compares favorably with previous proposals~\cite{li2007,spivak2009,li2010,cameron2014,durand2013,bradshaw2014,wang2014,hayat2015,alizadeh2015,durand2016,chen2016,dionne2016,zhang2017,acebal2017,ho2017,cao2018,dionne2018}
  and implementations \cite{hernandez2013,tkachenko2014,donato2015,tkachenko2014b,dionne2017,Schnoering2018,krevets2019,Nker2019}
of optical chiral resolution as it allows for the determination of 
the chiral parameter of each individual trapped nanoparticle via the rotation of the equilibrium position under the effect of a transverse drag force. However, this method has only been applied so far to coated nanoparticles. The case of homogeneous chiral nanoparticles is not only different from a physical point of view, as their light scattering properties are distinct from coated ones, but also is more relevant for a number of chiral systems, such as chiral plasmonic nanocrystals~\cite{fan2012}. Here we show that our method can be applied to micrometer sized homogeneous chiral spheres. In addition, we provide a detailed derivation of our theoretical approach, which is based on a generalization of the Mie-Debye theory of optical tweezers \cite{epl,Mazolli2003} to the case of chiral trapped particles.

This paper is organized as follows. In section II, we review the basic formalism of electromagnetic scattering by a chiral sphere  using the formalism of Debye potentials. In addition, the optical torque exerted on trapped chiral particles is discussed in detail.
 In Sec. III, we present our results and discuss how the chirality affects the condition for optical tweezing. Finally, in Sec. IV we summarise our findings and conclusions.
 

\section{Methodology}

\subsection{Electromagnetic fields in chiral media}

In a chiral medium, the electromagnetic fields ${\bf E}$ and ${\bf H}$ are coupled by a chirality parameter $\kappa$. The constitutive relations may be written as \cite{lindell,lakhtakia,chan2014} 
\begin{eqnarray}
&&\textbf{D}=\epsilon_0\epsilon \textbf{E} +i \kappa\sqrt{\epsilon_0\mu_0} \, {\bf H} \nonumber \\ 
&&\textbf{B}= -i \kappa\sqrt{\epsilon_0\mu_0}\,\textbf{E} +\mu \mu_0\textbf{H} \label{C5const}
\end{eqnarray}
where ${\bf D}$ and  $ {\bf B}$ are the electric displacement and the magnetic field in the chiral material,  respectively. 
Furthermore,   $\epsilon$, $\mu$ are the effective relative permittivity and permeability of the medium, and $\epsilon_0$, $\mu_0$ are the absolute permittivity and  permeability of vacuum.  

The Maxwell's equations for chiral media in the frequency domain can be compactly written in a matrix form with the help of the constitutive relations (\ref{C5const})  as  

\begin{gather}
  \nabla \cdot  \begin{bmatrix}  \textbf{E}\\  { \textbf{H}} \end{bmatrix} = 0
  \hspace{15pt} , \hspace{15pt}
  \nabla \! \times \! \begin{bmatrix}   \textbf{E}\\  { \textbf{H}} \end{bmatrix}
 = K \!
     \begin{bmatrix}
   \textbf{E} \\
   {\textbf{H}}  
   \end{bmatrix}
\end{gather}
where
$$
K=  \begin{bmatrix}
   k_0 \kappa & i k_0 \mu { \sqrt{\mu_0 / \epsilon_0}}  \\
   { - i k_0 \epsilon \sqrt{\epsilon_0 / \mu_0}}  & k_0 \kappa  
   \end{bmatrix}.
   $$
and $k_0 = \omega \sqrt{\mu_0 \epsilon_0} = \omega/c$. In addition, a direct inspection of (\ref{C5const}) shows that  the transverse character of {\bf E} and {\bf H} is only ensured if $\mu \epsilon \neq \kappa^2$, so we assume that such condition holds from now onwards \cite{footnote}.
  By introducing a linear transformation $A$, also known as Bohren decomposition, that mixes {\bf E} and {\bf H} \cite{bohren-huffman,lakhtakia} 
 \begin{gather}
  \begin{bmatrix}
   \bf{\Lambda_+} \\
   \bf{\Lambda_-}  
   \end{bmatrix}
 {\small = \frac{1}{2}
 \begin{bmatrix}
 \sqrt{\mu_0/\epsilon_0} &  i \sqrt{\mu/\epsilon} \\
  i \sqrt{\epsilon/\mu} & \sqrt{\epsilon_0/\mu_0} 
 \end{bmatrix} }
  \begin{bmatrix} \bf{E}\\   { \bf{H}} \end{bmatrix}
 = A
 \begin{bmatrix} \bf{E}\\   { \bf{H}} \end{bmatrix} \, ,
    \label{C5em in chiral}
\end{gather}
it is simple to show that we get uncoupled Helmholtz equations for each of the vectors ${\bf \Lambda}_{\sigma}, \; \sigma=\pm 1:$
\beq
&&\nabla ^{2}{\bf{\Lambda}_{\sigma}}+k_{\sigma} ^{2}\bf{\Lambda}_\sigma=0
\label{C5Helmholtz}
\eeq
where $k_{\sigma} = k_0 (\sqrt{\mu \epsilon}- \sigma \kappa).$
The parameter $\sigma$ represents the helicity of the propagating waves, as 
 $\bf{\Lambda}_+$ and $\bf{\Lambda}_-$  are left- and right-hand eigenmodes of incident circularly polarized waves that independently satisfy the Maxwell's equations in the frequency domain. 
 Finally, given the dispersion relation for $k_{\sigma}$, it is convenient to define a chiral refraction index $m_{\sigma}$
\be
m_{\sigma} = \sqrt{\mu \epsilon} -\sigma \kappa
\ee
that effectively characterizes the propagation of  circularly polarized waves of helicity $\sigma.$

\begin{figure}
\includegraphics[width = 3.in]{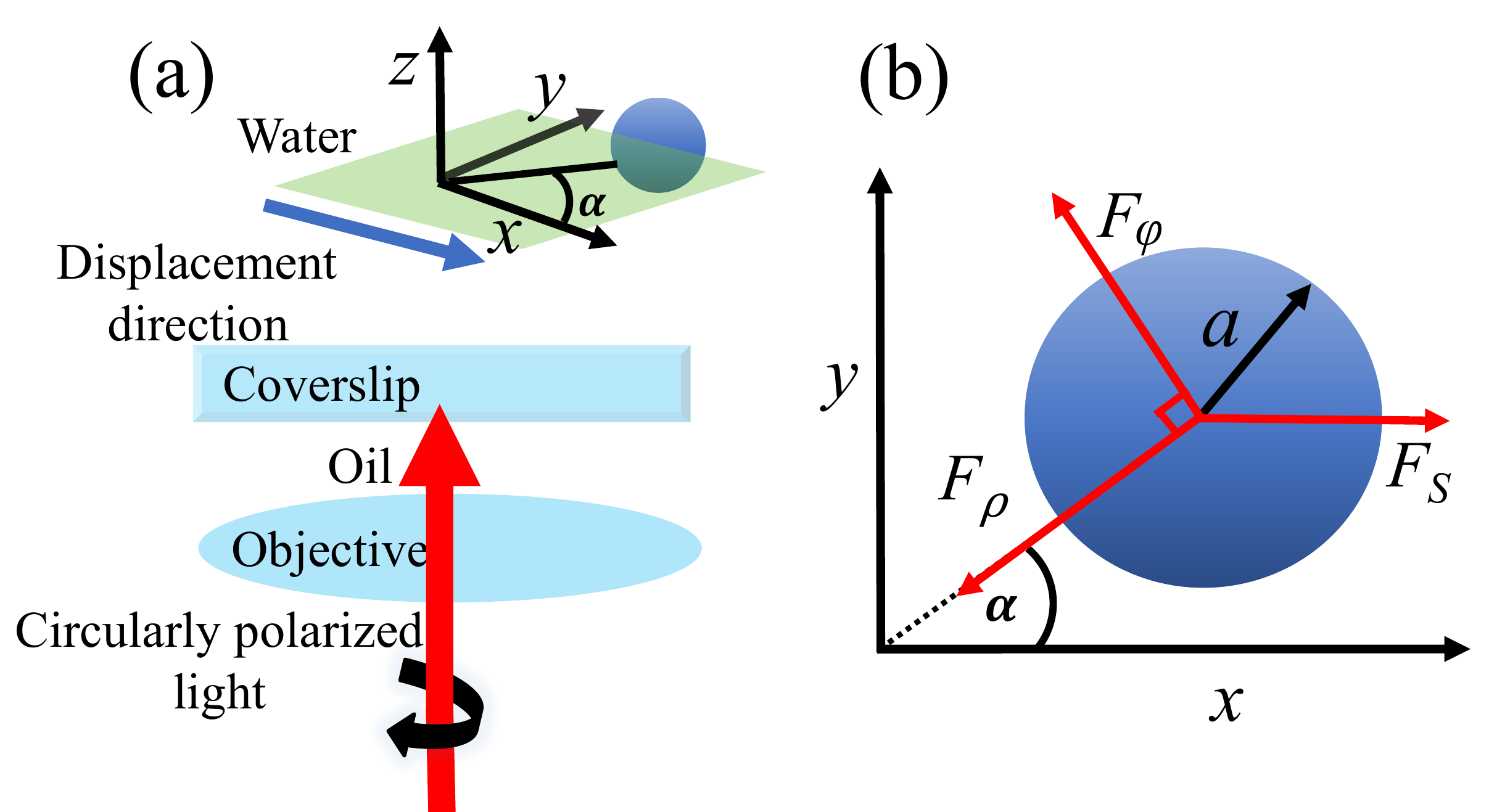}
\caption{ (a) Schematic representation of
enantioselective optical tweezing.
A single circularly-polarized  Gaussian laser beam is focused at the focal plane of the objective lens
in order to optically trap homogeneous chiral microspheres dispersed in aqueous solution. The sample is driven laterally at a constant speed. 
 (b) As the trapped sphere is laterally displaced by the resulting Stoles drag force $F_S$, its equilibrium position rotates around the optical $z$ axis
  by an angle $\alpha.$ The cylindrical optical force components $F_{\rho}$ and  $F_{\phi}$ are shown.  
} 
 \label{C5setup}
\end{figure}

\subsection{Mie series solution for a chiral sphere}
\label{C5Section_Mie}

The solution of the Helmholtz equation (\ref{C5Helmholtz}) in spherical coordinates is thoroughly discussed in the literature \cite{bohren-huffman,Born-Wolf}, but the chiral case presents a non-trivial additional element: while propagating eigenmodes correspond to the vector fields  ${\bf \Lambda}_{\sigma}$, 
 the boundary conditions involve the fields ${\bf E, D, B} $ and $ {\bf H}$. 
 Fortunately, the linear relation (\ref{C5em in chiral}) allows a straightforward solution 
 for the scattered field by a chiral sphere (given an incident plane wave)
 \cite{bohren-huffman,lakhtakia,bohren}:
\begin{equation}
\begin{aligned}
{\bf E}_{s}= E_0 \sum^{\infty}_{\ell} (i)^{{\ell}}\frac{2{\ell}+1}{{\ell}({\ell}+1)}  (i a_\ell N_{e1{\ell}}^3\\-b_{\ell}M_{o1{\ell}}^3 + c_{\ell}M_{e1{\ell}}^3 - id_{\ell}N_{o1{\ell}}^3) \\
{\bf H}_{s}= \frac{k}{\omega \mu} E_0 \sum^{\infty}_{\ell} (i)^{{\ell}}\frac{2{\ell}+1}{{\ell}({\ell}+1)}
( a_{\ell}M_{e1{\ell}}^3\\+ ib_{\ell}N_{o 1{\ell}}^3 - ic_\ell N_{e1{\ell}}^3 - d_{\ell}M_{o 1{\ell}}^3),
\label{C5SCA}
\end{aligned}
\end{equation} 
where $M_{e1\ell}^3$, $M_{o1\ell}^3$, $N_{e1\ell}^3$  and $N_{o 1 \ell}^3$ are the vector spherical harmonics \cite{bohren-huffman}. The scattering amplitudes   $ a_{\ell} $,  $ b_{\ell} $,  $ c_{\ell}$  and  $ d_{\ell} $  are commonly known as Mie  coefficients  \cite{bohren-huffman,lakhtakia,bohren,shang,chan2014}
 
\begin{eqnarray}
&&a_{\ell} = \frac{V_{\ell}(-)A_{\ell}(+)+V_{\ell}(+)A_{\ell}(-)}{W_{\ell}(+)V_{\ell}(-)+V_{\ell}(+)W_{\ell}(-)} \nonumber \\
&&b_{\ell}= \frac{W_{\ell}(+)B_{\ell}(-)+W_{\ell}(-)B_{\ell}(+)}{W_{\ell}(+)V_{\ell}(-)+V_{\ell}(+)W_{\ell}(-)} \nonumber \\
&&c_{\ell} =-d_{\ell} =  i\frac{W_{\ell}(-)A_{\ell}(+)-W_{\ell}(+)A_{\ell}(-)}{W_{\ell}(+)V_{\ell}(-)+V_{\ell}(+)W_{\ell}(-)}
 \label{C5MC}
\end{eqnarray}
with 
\begin{equation}
\begin{aligned}
W_{\ell}(\sigma) = m\psi_{\ell}(y_{\sigma}) \xi_{\ell}^{'}(x)-\xi_{\ell}(x)\psi_{\ell}^{'}(y_{\sigma})\\
V_{\ell}(\sigma) = \psi_{\ell}(y_{\sigma}) \xi_{\ell}^{'}(x)-m \xi_{\ell}(x)\psi_{\ell}^{'}(y_{\sigma})\\
A_{\ell}(\sigma) = m\psi_{\ell}(y_{\sigma}) \psi_{\ell}^{'}(x)-\psi_{\ell}(x)\psi_{\ell}^{'}(y_{\sigma})\\
B_{\ell}(\sigma ) = \psi_{\ell}(y_{\sigma}) \psi_{\ell}^{'}(x)-m \psi_{\ell}(x)\psi_{\ell}^{'}(y_{\sigma})
\end{aligned}
\end{equation}
where $m = m_+ m_- /2(m_+ + m_-).$ The Riccati-Bessel functions  \cite{Abramowitz} $\psi_{\ell}$, $\xi_{\ell}$  are evaluated either at the 
size parameter $x=\sqrt{\epsilon_w}k_0 a$ defined with respect to the wavelength in the non-magnetic achiral host medium (relative electric permittivity $\epsilon_w$)  
or at $y_{\sigma}=m_{\sigma}x/\sqrt{\epsilon_w}.$
Finally, for future convenience, we introduce the  Debye potentials $\Pi^{E}$ and $\Pi^{M}$ for electric and magnetic multipoles \cite{Born-Wolf} 
\begin{equation}
\begin{aligned}
\Pi^{E} ({\bf r})= \sum^{\infty}_{\ell}\frac{({\bf r}\cdot {\bf E})_{}}{{\ell}({\ell}+1)}\\
\Pi^{M}({\bf r}) = \sum^{\infty}_{\ell}\frac{({\bf r}\cdot {\bf H})}{{\ell}({\ell}+1)} \, ,
\label{C5DB}
\end{aligned}
\end{equation}
which satisfy a (scalar) Helmholtz equation. Using Eq. (\ref{C5DB})  we write the electromagnetic fields in terms of the scalar Debye potentials as follows:
\begin{equation}
 \begin{aligned}
{\bf E}=  \nabla \times \nabla \times \left({\bf r}\, \Pi^{E}\right) + i\omega \mu \,  \nabla \times\left({\bf r}\, \Pi^{M}\right)\\
{\bf H}=  \nabla \times \nabla \times\left({\bf r}\, \Pi^{M}\right) - i\omega \epsilon\,  \nabla \times\left({\bf r}\, \Pi^{E}\right) \, .
\end{aligned} \label{C5Field_potentials}
\end{equation} 

 \subsection{Mie-Debye theory of optical tweezers for chiral sphere}
 
 As it is depicted in Fig. 1, an optical tweezers setup consists of a laser beam strongly focused by a high numerical aperture (NA) objective. Stable trapping in the focal region is possible for certain ``goldilocks'' combinations of the optical properties of the object to be trapped and the surrounding medium.
 We consider a $1064 \, \rm nm$ Gaussian beam, circularly polarized with helicity $\sigma$, at the objective 
entrance port. The beam is focused by the objective 
 into the sample region containing an aqueous suspension (refractive index $n_w=\sqrt{\epsilon_w}$) of 
 homogeneous chiral spheres, with effective relative permittivity $\epsilon_s$ and chirality $\kappa$. 
The focused beam in the sample region is represented in terms of the so called  Richards-Wolf integral representation \cite{RichardsWolf,Mazolli2003}
\beq
\mathbf{E}_{\rm in}(\mathbf{r}) &&\hspace{-10pt}= {E}_{0}\int_{0}^{2\pi}d\varphi_k\int_{0}^{\theta_{0}}d{\theta_k}\sin \theta_k \sqrt{\cos \theta_k} \,
e^{-\gamma_f^2\sin^2\theta_k} \nonumber \\
 && \hspace{60pt} \times \, e^{i\mathbf{k}(\theta_k,\varphi_k)\cdot(\mathbf{r}+\mathbf{r}_s)}\mathbf{\hat{\epsilon}_{\sigma}^{\prime}}(\theta_k, \varphi_k),
\label{C5RW}
\eeq
where 
{we set the origin at the center of the sphere and} the focal point is at position $-\mathbf{r}_s$. 
The region of integration in Fourier space is defined by the angle $\theta_0,$ which in turn depends on the
objective numerical aperture {\rm NA}:
$\sin\theta_0=\min\{1,\mbox{NA}/n_w\}.$
 $\gamma_f$ is the ratio of the objective focal length to the laser beam waist at the objective entrance port and 
  $\mathbf{\hat{\epsilon}_{\sigma}^{\prime}}(\theta_k, \varphi_k)=\mathbf{\hat{x}^{\prime}}(\theta_k, \varphi_k)+i\sigma
 \mathbf{\hat{y}^{\prime}}(\theta_k, \varphi_k).$ The  unit vectors $\mathbf{\hat{x}^{\prime}}(\theta_k, \varphi_k)$ and $\mathbf{\hat{y}^{\prime}}(\theta_k, \varphi_k)$ are obtained from $\mathbf{\hat{x}}$ and  $\mathbf{\hat{y}}$ by rotation with Euler angles $\alpha=\varphi_k,$ $\beta = \theta_k$ and $\gamma=-\varphi_k$.
 We have assumed an ideal aplanatic optical system when writing Eq.~(\ref{C5RW}). 
 Optical aberrations can be included as additional phase factors in the integrand in the r.-h.-s. of
  (\ref{C5RW})  \cite{Dutra2014} and the resulting optical force can be derived by following the steps presented below. 
 
We take $\mathbf{E}_{\rm in}(\mathbf{r})$ as the incident field on the chiral microsphere. 
 The strategy is to take each plane wave component of Eq.  (\ref{C5RW}), solve the scattering problem using the results of section \ref{C5Section_Mie}, and superpose the solutions appropriately. There is, however, a caveat: the expressions in last section assume a $\hat{\bf z}$-propagating incident wave, and in (\ref{C5RW}) we have components propagating in many different directions, so we must rotate our coordinate system in order to have the solution for an arbitrary incident wave. This is accomplished by the matrix elements of finite rotations $d^j_{m, m'}(\theta)$\cite{Edmonds}.  After these operations, the Debye potentials associated to an incident circular polarized plane wave (helicity $\sigma$) propagating along an arbitrary direction defined by the spherical angles $(\theta_k,\phi_k)$ are given by
\begin{equation}
\begin{aligned}
 \Pi^{E, {\rm in}}_{{\bf k}, \sigma} (r,\theta,\phi) = \sigma \frac{i E_0}{k} \sum_{{\ell}} i^{{\ell}}   j_{\ell}{(kr}) \sqrt{\frac{4\pi(2{\ell}+1)}{{\ell}({\ell}+1)}}
 Y^{(\sigma)}_{\ell}(\theta ,\phi)\\
 \Pi^{M, {\rm in}}_{{\bf k}, \sigma} (r,\theta,\phi) = \frac{E_0}{k} \sum_{{\ell}} i^{{\ell}} j_{\ell}{(kr}) \sqrt{\frac{4\pi(2{\ell}+1)}{{\ell}({\ell}+1)}} 
Y^{(\sigma)}_{\ell}(\theta ,\phi) \, ,
\end{aligned}
\end{equation} 
where $j_{\ell}(x)$ are the spherical Bessel functions \cite{Abramowitz} and
$$
Y^{(\sigma)}_{\ell}(\theta,\phi) := \sum^{\ell}_{{m}=-\ell} Y_{\ell{m}}(\theta,\phi) e^{-i \phi_k(m-\sigma)} d^{\ell}_{m ,{\sigma}}(\theta_k) \, .
$$
Solving the scattering problem for general $\theta_k$ and $\phi_k$, superposing the solutions according to (\ref{C5RW}), {and undertaking a translation to a coordinate system centered at the focal point --- so that the sphere's center is finally at position $\mathbf{r}_s$ ---, we arrive} at the Debye potentials for the total fields
\begin{eqnarray}
&&\hspace{-20pt}\Pi_{\sigma}^{E, \rm{tot}}=   \sigma \frac{i E_0}{k} \sum_{\ell,m} i^{\ell}\gamma_{\ell,m}^{\sigma} \left [j_{\ell}{(kr})-(a_{\ell}+i\sigma  d_{\ell}) h_{\ell}(kr)\right]  \times \nonumber \\ 
&& \hspace{60pt} Y_{\ell m} (\theta ,\phi)  e^{-i(m-\sigma)\phi_s} \\ \nonumber \\
&&\hspace{-20pt}\Pi_{\sigma}^{M, \rm{tot}}=   \frac{E_0}{k} \sum_{\ell,m} i^{\ell} \gamma_{\ell,m}^{\sigma}  \left [  j_{\ell}{(kr})-(b_{\ell} -i \sigma  c_{\ell} )h_{\ell}(kr)\right]  \times \nonumber \\ 
&&\hspace{60pt}   Y_{\ell m} (\theta ,\phi)  e^{-i(m-\sigma )\phi_s} \, ,
\label{C5Debye_potentials_tot}
\end{eqnarray} 
where $h_{\ell}(kr)$ is the spherical Hankel function of the first kind. The coefficients 
\begin{eqnarray}
&&\hspace{-20pt}\gamma_{\ell,m}^{\sigma}= 2\pi \sqrt{\frac{4\pi(2{\ell}+1)}{{\ell}({\ell}+1)}}(-i)^{m-\sigma} \int_{0}^{\theta_{0}}d\theta_{k}\sin\theta_{k} \times \nonumber \\
&& \sqrt{\cos\theta_{k}} d^\ell _{m,\sigma}(\theta_{k})J_{m-\sigma}(k\rho_s\sin\theta_{k})e^{ikz_s\cos{\theta_{k}}} 
\end{eqnarray}
are written in terms of the cylindrical Bessel functions $J_{m-\sigma}$ of integer order $m-\sigma.$

Here we have expressed the microsphere position ${\bf r}_s=(\rho_s,\varphi_s,z_s)$ in cylindrical coordinates. 
When taking $\kappa$ to zero ($c_\ell=d_\ell=0$) we recover the Debye potentials for an achiral sphere \cite{Mazolli2003}. 
However, the most interesting property of Eqs.~(\ref{C5Debye_potentials_tot}) is that they have the same form as the Debye potentials of an achiral sphere, provided the following replacements are made
%
\begin{eqnarray}
&& A_\ell \leftrightarrow a_\ell+ \,  i \,\sigma d_\ell\label{equivalence} \\
&& B_\ell \leftrightarrow b_\ell -  i\sigma  \, c_\ell \, ,
\label{equivalence2}
\end{eqnarray}
where $A_\ell$ and $B_\ell$ would be the Mie coefficients of the equivalent achiral sphere scattering a circularly polarized beam of helicity $\sigma.$ Equations (\ref{equivalence}) and (\ref{equivalence2}) show that one can map the problem of light scattering by a homogeneous chiral sphere into the simpler original Mie scattering problem.

As the final step in the derivation, we compute the optical force upon the sphere by integration of the Maxwell stress tensor:
\begin{equation}
{\bf F}= \lim_{r\rightarrow \infty}\left[ -\frac{r}{2} \int_{S} {\bf r}\left( \epsilon_w \epsilon_0 E_{\rm tot}^2 + \mu_0 H_{\rm tot}^2 \right) d\Omega  \right] 
\label{C5Max}
\end{equation}
where ${\bf E}_{\rm tot}$ and ${\bf B}_{\rm tot}$ are the total electric and magnetic fields 
and the Gaussian spherical surface $S$ is taken at infinity. At this point it is useful to define the (dimensionless) efficiency factor ${\bf Q}$ 
\begin{equation}
{\bf Q} = \frac{c}{n_w P} {\bf F} \, ,
\end{equation}
where ${\bf F}$ is given by (\ref{C5Max}), $P$ is the incident laser power  at the sample region and $c$ is speed of light. 
The efficiency may be conveniently split into extinction and  scattering terms
\begin{equation}
{\bf Q} = {\bf Q}_{e} + {\bf Q}_{s},  \label{C5Q}
\end{equation}
where the extinction term ${\bf Q}_e$ is connected to the rate of momentum removal from the incident beam. 
Part of this momentum ends up in the scattered field at a (normalized) rate $-{\bf Q}_s$, so the total (normalized) momentum transfer rate to the particle is ${\bf Q}_e + {\bf Q}_s$. After a long but straightforward calculation \cite{epl, Mazolli2003}, we get the expressions for the efficiency components $Q_z$, $Q_s$ and $Q_{\phi}$, related to the axial, radial and azimuthal force components, respectively. Explicitly, we have 

\begin{widetext}

\begin{itemize}

\item Extinction components
\begin{equation}
Q_{ez}(\rho_s,z_s)=\frac{4\gamma_f^2}{AN}{\rm Re}\sum_{\ell m}(2\ell+1)
G^{(\sigma)}_{\ell,m}
(A_{\ell}+B_{\ell})
G'^{(\sigma)*}_{\ell,m},
\label{C5Qez}
\end{equation}

\begin{equation}
Q_{e\rho}(\rho_s,z_s)=\frac{2\gamma_f^2} {AN}{\rm Im}\sum_{\ell m}(2\ell+1)G^{(\sigma)}_{\ell,m}
(A_{\ell}+B_{\ell}) 
\left(G^{(\sigma)-}_{\ell ,m+1} - G^{(\sigma)+}_{\ell,m-1}\right)^*
 \label{C5Qerho}
\end{equation}

\begin{equation}
Q_{e\phi}(\rho_s,z_s)=-\frac{2\gamma_f^2} {AN}{\rm Re}\sum_{\ell m}(2\ell+1)G^{(\sigma)}_{\ell,m}
(A_{\ell}+B_{\ell}) 
\left(G^{(\sigma)+}_{\ell,m-1}+G^{(\sigma)-}_{\ell,m+1}\right)^*
 \label{C5Qerho}
\end{equation}

\item Scattering components
\begin{eqnarray}
Q_{s z}(\rho_s,z_s) && \hspace{-10pt}= -\frac{8\gamma_f^2}{AN}{\rm Re}\sum_{\ell m}\frac{\sqrt{\ell(\ell+2)(\ell+m+1)(\ell-m+1)}}{\ell+1}
  (A_{\ell}A_{\ell+1}^{*}+B_{\ell}B_{\ell+1}^{*})
G^{(\sigma)}_{\ell,m}G^{(\sigma)*}_{\ell+1,m} \nonumber \\ 
 &&-\frac{8\gamma_f^2}{AN}\, \sigma \,{\rm Re}\sum_{\ell m}
\frac{(2\ell+1)}{\ell(\ell+1)} m A_{\ell}B_{\ell}^{*}\vert G^{(\sigma)}_{\ell,m}\vert^2
, \label{C5Qszp}
\end{eqnarray} 

\begin{eqnarray}  
Q_{s\rho}(\rho_s,z_s) = \frac{4\gamma_f^2}{AN}\sum_{\ell m}\frac{\sqrt{\ell(\ell+2)(\ell+m+1)(\ell+m+2)}}{\ell+1}
 {\rm Im}\biggl\lbrace (A_{\ell}A_{\ell+1}^{*}+B_{\ell}B_{\ell+1}^{*}) 
\left[ G^{(\sigma)}_{\ell,m}G^{(\sigma)*}_{\ell+1,m+1} 
\right. \nonumber \\
 \left.
  +G^{(\sigma)}_{\ell,-m}G^{(\sigma)*}_{\ell+1,-(m+1)}\right]\biggl\rbrace
 -\frac{8\gamma_f^2}{AN}\sigma\sum_{\ell m} \frac{(2\ell+1)}{\ell(\ell+1)}\sqrt{(\ell-m)(\ell+m+1)}   
 \; {\rm Re}(A_{\ell}B_{\ell}^{*})
\, {\rm Im}( G^{(\sigma)}_{\ell,m}G^{(\sigma)*}_{\ell,m+1})
     \label{C5Qsrho}
\end{eqnarray} 

\begin{eqnarray}  
Q_{s\phi}(\rho_s,z_s) = -\frac{4\gamma_f^2}{AN}\sum_{\ell m}\frac{\sqrt{\ell(\ell+2)(\ell+m+1)(\ell+m+2)}}{\ell+1}
 {\rm Re}\biggl\lbrace (A_{\ell}A_{\ell+1}^{*}+B_{\ell}B_{\ell+1}^{*})
\left[ G^{(\sigma)}_{\ell,m}G^{(\sigma)*}_{\ell+1,m+1} \right. \nonumber \\
 \left.-G^{(\sigma)}_{\ell,-m}G^{(\sigma)*}_{\ell+1,-(m+1)}\right]\biggr\rbrace
+\frac{8\gamma_f^2}{AN}\sigma\sum_{\ell m}
 \frac{(2\ell+1)}{\ell(\ell+1)}\sqrt{(\ell-m)(\ell+m+1)} \; {\rm Re}(A_{\ell}B_{\ell}^{*})
\,{\rm Re}( G^{(\sigma)}_{\ell,m}G^{(\sigma)*}_{\ell,m+1}) \, ,
   \label{C5Qsphi}
\end{eqnarray} 

\end{itemize}

\end{widetext}

 We have defined 
\(
A = 1-e^{-2\gamma_f^2\,\sin^2\theta_0}
\)
as the fraction
of the trapping beam power that fills the objective entrance aperture and propagates into the sample. 
Explicit expressions for
the multipole coefficients $G_{\ell m}^{(\sigma)}$, ${G'}_{\ell m}^{(\sigma)}$ and $G_{\ell m}^{(\sigma)\pm}$  are given in the appendix.


\section{Results and Discussion}

 In this section, we present our numerical results for the optical forces and optical torques in the framework of the MDSA formalism derived in Sec. ~II. In all the following examples we take the refractive indexes of the sphere and aqueous suspension  to be $n_s = 1.58$ and $n_w = 1.332,$ 
  respectively. The vacuum wavelength of the laser beam is $\lambda_0=2\pi/k_0=1064\, \rm nm$, and the other relevant parameters are
  $\gamma_f=1.226$  and NA$=1.4$.

\subsection{Optical trapping}

\begin{figure}
\includegraphics[width = 3.4 in]{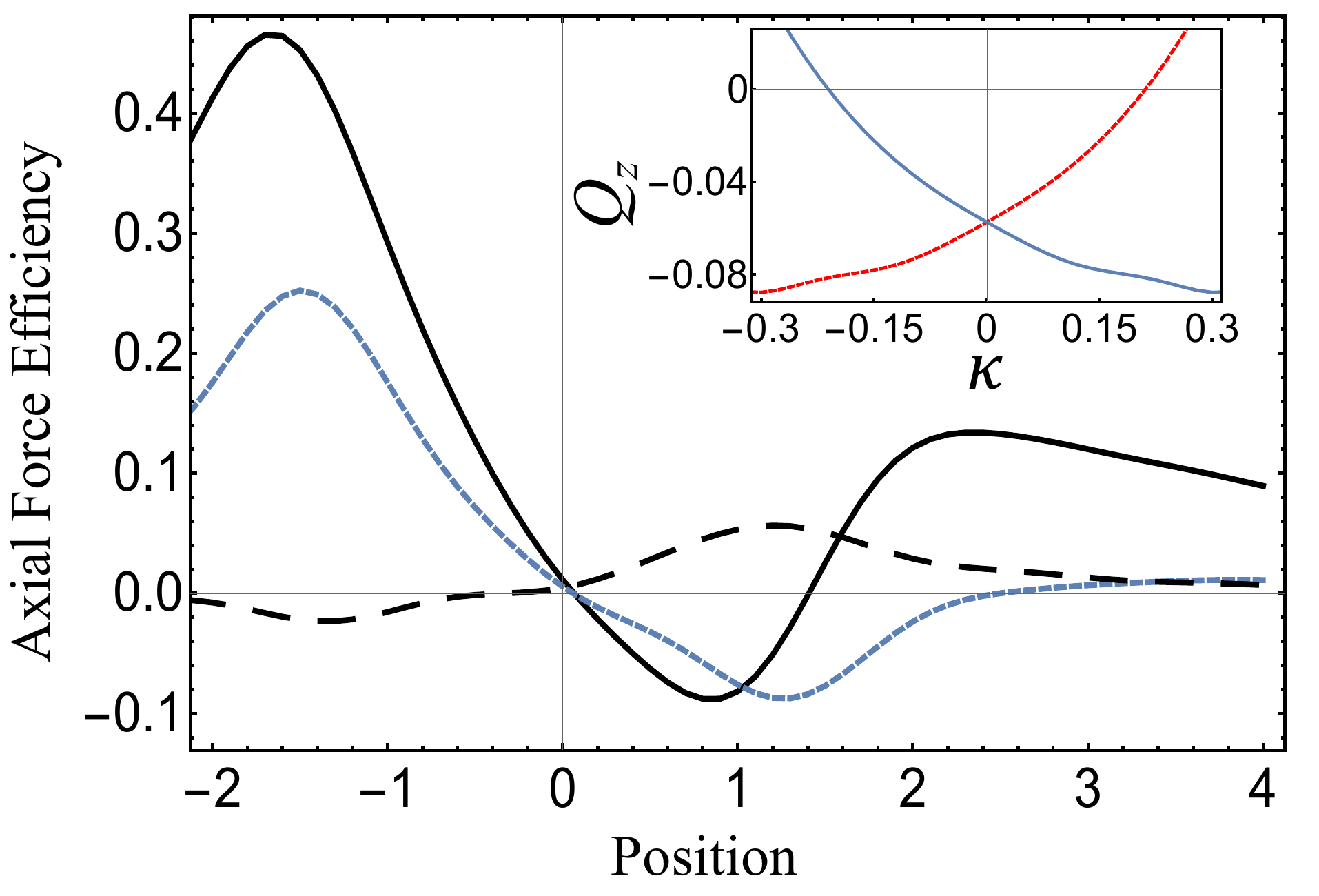}
\caption{ Dimensionless axial force efficiency $Q_z$ as a function of position (in units of sphere radius) along the laser axis for different values of the 
chirality parameter: $\kappa=-0.3$ (black line), $\kappa=0$ (dotted line) and $\kappa=0.3$ (dashed line). The  incident beam is 
right-circularly polarized ($\sigma=-1$) and the microsphere radius is  $ a=1\, \mu {\rm m}$.
Inset: $Q_z$ as a function of  $\kappa$ at a fixed axial position  $z/a=1.2$  for  $\sigma=-1 $ (red dotted line) and  $\sigma=1$  (blue line). }
  \label{C5a1}
\end{figure}

Due to axial symmetry, it is sufficient to analyze the optical force $F_z$ to ensure 3-D trapping stability of optical trap. In Fig. \ref{C5a1},  we calculate  the normalized  axial force acting on a microsphere as a function of sphere position along the laser axis  
when taking 
 right-circular polarization (RCP), which corresponds to helicity $\sigma=-1.$.
We clearly see  an enantioselection effect  taking place. For instance, the RCP beam attracts the right-handed chiral particles with $\kappa=-0.3$ (black line) towards the paraxial focal plane with stable equilibrium position close to the focus  and  repels the left-handed chiral enantiomer with $\kappa=0.3 $ out of the illuminated area (dashed line).
Due to mirror symmetry, this result is exactly reversed for left circularly polarized (LCP) beams, so that in this case the spheres with positive chirality get trapped.  This is illustrated by the inset of Fig. \ref{C5a1}, where we plot the normalized axial force as a function of chirality for left and right polarization of the beam. Therefore, in a racemic mixture one could potentially use this effect to trap exclusively right- or left-handed particles by using a RCP or a LCP beam. 
In addition, from the inset of  Fig. \ref{C5a1} it is also clear that despite favoring ``same sign'' trapping (RCP traps right-handed particles and vice versa), ``oposite sign'' trapping is also possible up to a critical value $|\kappa_0| \approx 0.22$. 

\begin{figure}
\includegraphics[width = 3.4 in]{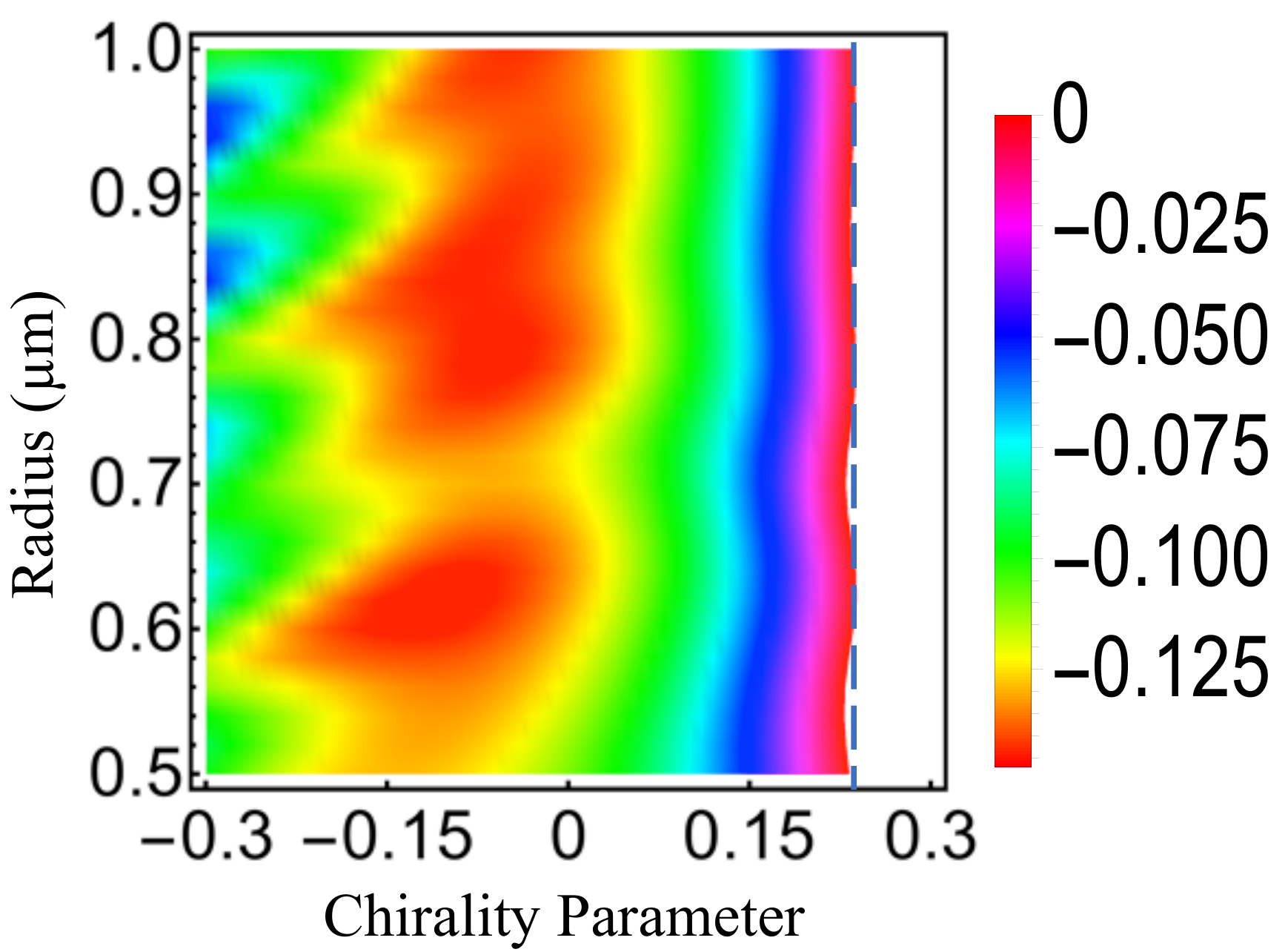}
\caption{ Density plot of the dimensionless axial force efficiency $Q_z$ at axial position  $z/a=1.2$.
The horizontal and vertical axes represent the chirality parameter and the microsphere radius, respectively.
Only negative values of $Q_z$ are represented by the
 color scale. The vertical line indicates the critical value
  $\kappa_0$ above which no trapping takes place as the optical force becomes positive.}
 \label{C5a3}
\end{figure}

We analyse more thoroughly the interplay of chirality and particle size in Fig. \ref{C5a3}.  The colored area shows the negative optical force (trapping force) on right handed  chiral enantiomers for RCP incident light. The most important feature of this plot is the weak dependence of the force upon the sphere radius, at least for the range considered. This means that, at least for sufficiently large  values of $\kappa$, the enantioselection effect is quite robust against variation of the radius, and could be effective even in solutions where the spheres size dispersion is large.
It is important to emphasize that such values of $\kappa$ are within the reach of state-of-the-art plasmonic and metamaterial chiral nanoparticles~\cite{dionne2016,fan2012,wang2014}.

\subsection{Optical torque}

In the preceding section,  we have shown that one can perform chiral resolution of a racemic mixture by 
selective trapping of chiral enantiomers that match the handedness of the laser beam.
In this section, we discuss another  method that is  based on the transfer of spin angular momentum (SAM) from the focused beam to the trapped microsphere, allowing to determine the chiral parameter of single particles. 

In order to exploit the SAM transfer, first we stably trap a microsphere and then produce a water flow by  driving
the sample laterally
 (say, along the $x-$direction), as indicated in  Fig.~\ref{C5setup}. A drag force  $F_S$ is then generated along the $x-$direction, that displaces the sphere away from the center of the optical trap and gives rise to a restoring force $F_\rho$. In addition, when the trapping beam is circularly polarized, it is intuitive that by pushing the sphere away from the beam axis we should also get a force along the azimuthal direction $F_\phi$,  as illustrated in Fig.~\ref{C5setup}. This azimuthal force component
 essentially converts (some of) the spin angular momentum of the beam to angular momentum of the sphere, until it equilibrates at an angle $\alpha = \arctan(F_{\phi} / F_{\rho})$. Explicit expressions for such force components were given in Sec. II.
 It is more than reasonable to expect that such equilibrium angle should depend considerably upon the sphere chirality, and that is precisely what we show below.

\begin{figure}
\includegraphics[width = 3. in]{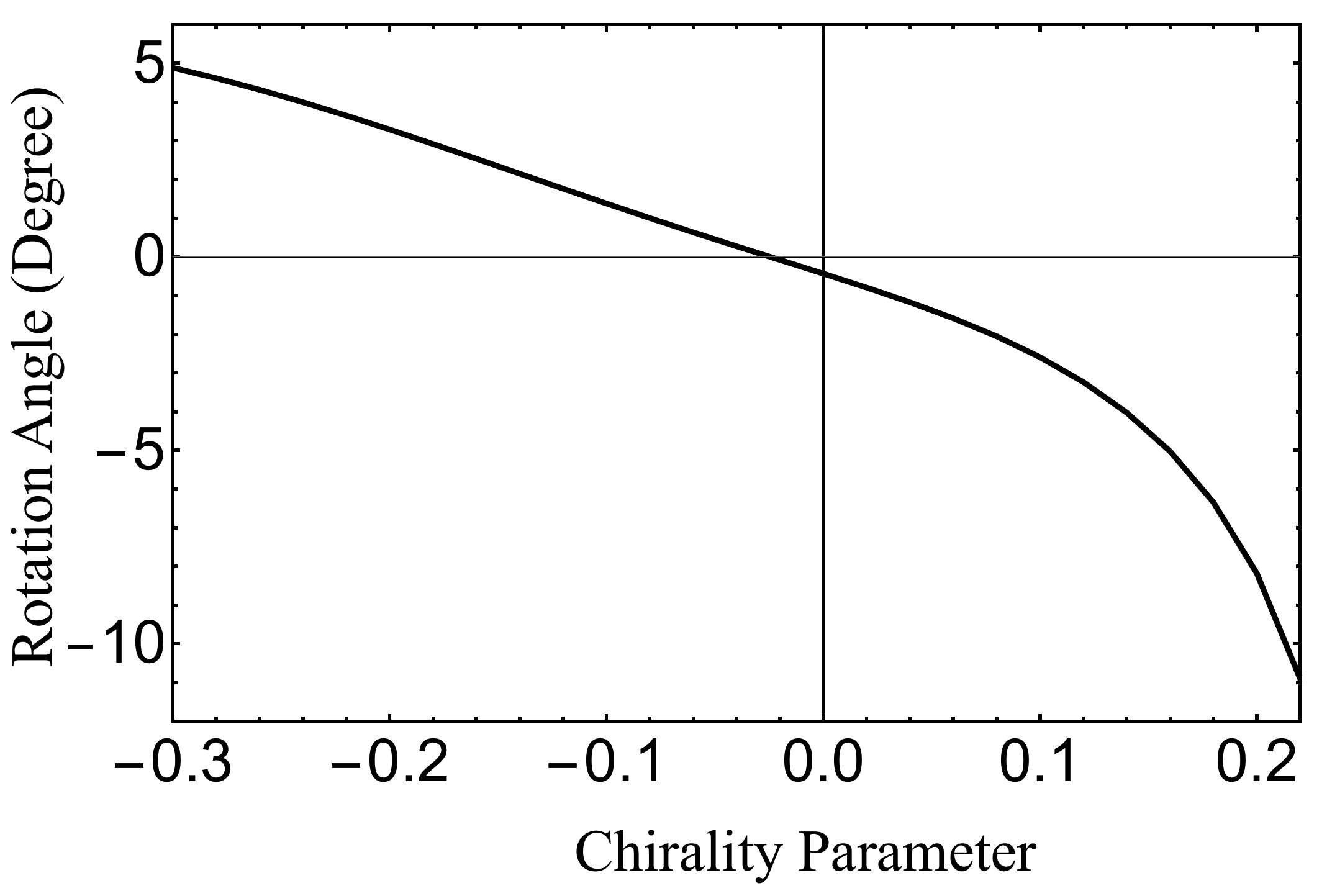}
\caption{Rotation angle $\alpha$ of a chiral microsphere of radius  {$a=0.6\, \mu{\rm m}$} as a function of the chirality parameter $\kappa.$ 
The incident beam is right-circularly polarized ($\sigma=-1$). }
 \label{C5a4}
\end{figure}

Fig. \ref{C5a4} shows the optical torque exerted on the sphere in terms of the rotation angle $ \alpha $ in degrees for a RCP trapping beam ($\sigma=-1$).
 First of all, it is important to mention that for a small decrease in the chirality parameter of magnitude $ \delta \kappa = 10^{-2}$  the sphere undergoes a measurable additional 
 rotation of $\delta \alpha=0.2-0.5$ degrees (with the largest variation for the highest positive values of $\kappa$ shown in the figure). 
 Such strong variation of the rotation angle suggests a suitable scheme to determine and characterize the chiral parameter  of a single particle, which remains a challenging task so far~\cite{dionnereply}.

When the rotation angle becomes positive for negative values of $\kappa$
in Fig.~\ref{C5a4}, it means that the scattered field carries a magnitude of angular momentum
that exceeds that of the 
incident field \cite{kaina2019}, so that conservation of angular momentum leads to a torque opposite to the input spin. Such negative torque corresponds to a positive rotation angle $\alpha$ because we take $\sigma=-1$ here. 
 By increasing the value of $\kappa$, the rotation angle not only changes its sign but also achieves very large values in comparison to the case of  achiral homogenous spheres treated so far~\cite{kaina2019}, showing that chirality leads to a significant enhancement of
 the light-particle angular momentum transfer. 

 Figure \ref{C5a4} also shows that  $\alpha$ changes its sign at a relative small value, $\kappa\approx  -0.02$. 
 This result demonstrates that enantioselection can also be attained by detecting the sign of the rotation angle of a particle trapped using optical tweezers, in addition to the previous chiral resolution mechanism based on selective trapping of particles of a given handedness, as previously presented.    

\begin{figure}
\includegraphics[width = 3.4 in]{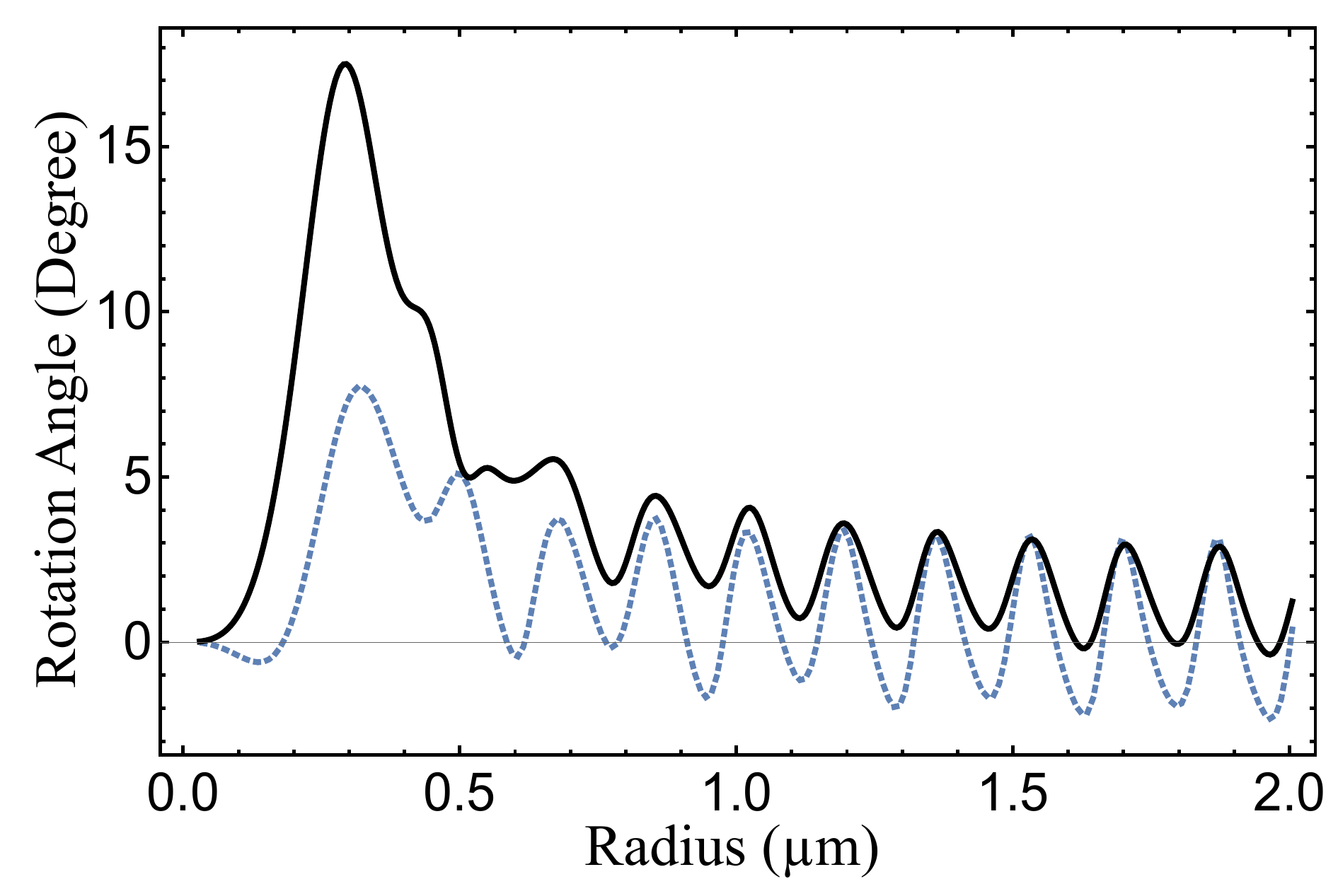}
\caption{Microsphere rotation angle versus sphere radius. The solid and dotted lines correspond to
   $\kappa= -0.3$ and  $\kappa= 0,$ respectively.  The incident beam is right-circularly polarized ($\sigma=-1$). }
 \label{C5a5}
\end{figure}

In Fig.~\ref{C5a5}, we plot the variation of the rotation angle $\alpha$ versus
the microsphere radius for two fixed values of the chirality parameter: $\kappa= -0.3$ (solid) and  $\kappa= 0$ (dotted). 
As in Fig.~\ref{C5a4}, we take RCP corresponding to $\sigma=-1.$
For small microspheres
  the rotation angle is negative for
  achiral particles (positive torque)
  and positive for $\kappa=-0.3$ (negative torque). 
  In the Rayleigh scattering regime, $a \ll \lambda_0$, scattering is weak and both force and torque are dominated by the extinction contribution. 
  Indeed, very small particles behave as local probes of the input angular momentum, thus explaining the negative angles in the case of small achiral spheres.
  On the other hand, Fig.~\ref{C5a5} shows that the particle chirality strongly affects the sign of the optical torque even in the Rayleigh limit. 
  
  As the size parameter increases light scattering becomes more prominent and the transfer of angular momentum is amplified~\cite{SBradshaw2017}, leading to  much larger angles as illustrated by Fig.~\ref{C5a5}. 
  In particular, we find a peak value around $a\approx 0.3\,\mu{\rm m}$ of approximately $17^{\rm o}.$  Such a large rotation angle cannot be obtained with achiral microspheres~\cite{kaina2019}. As the radius is further increased into the ray optics range, the rotation angle for the achiral sphere 
  oscillates around zero, which is compatible with the vanishing optical torque in the ray optics approximation \cite{Mazolli2003}, whereas the curve for the chiral 
  particle suggests a nonzero torque in this limit.

\section{Conclusions}

In this paper we developed the theory of optical tweezing for optically active homogeneous microspheres. First we developed the Bohren decomposition in terms of Debye potentials representing the incident and scattered field. 
Then we used the Maxwell stress tensor to calculate the optical forces exerted on chiral spheres.  
Our numerical analysis predicts that circularly polarized light exerts different optical forces on left- and right-handed chiral microspheres, allowing for enantioselection.
We also have shown that angular momentum transfer is different for chiral microspheres of opposite handedness,  which not only greatly enhances optical rotation of the microsphere with respect to the achiral case treated so far but also leads to another chiral resolution method based on enantioselective optical torque. Altogether our findings pave the way for alternative chiral resolution methods based on optical tweezing and characterisation of chirality for the important class of homogeneous chiral microspheres, each and every one with its individual, unique chiral response.

\begin{acknowledgments}
We thank D. S. Ether jr and N. B. Viana for inspiring discussions.  This work has been supported by 
 the Brazilian agencies National Council for Scientific and Technological Development (CNPq), 
  Coordination for the Improvement of Higher Education Personnel (CAPES),  the National Institute of Science and Technology Complex Fluids (INCT-FCx),
and the Research Foundations of the States of Minas Gerais (FAPEMIG), Rio de Janeiro (FAPERJ) and S\~ao Paulo (FAPESP).

\end{acknowledgments}

\appendix

\section{ Multipole coefficients of the focused beam }

The multipole coefficients in expressions (\ref{C5Qez})-(\ref{C5Qsphi}) are given by 

\begin{widetext}

\begin{equation}\label{C5Gjm}
G_{\ell m}^{(\sigma)}({ \rho_s,z_s})=\int_{0}^{\theta_0}d\theta\,\sin\theta\sqrt{\cos {\theta}}\, e^{-\gamma_f^2\sin^2\theta}
d_{m,\sigma}^{\ell}(\theta)\, J_{m-\sigma}\left( k \rho_s \sin\theta\right) e^{i\,k\cos\theta\, z_s },
\end{equation}

\begin{equation}
G'^{(\sigma)}_{\ell,m}({ \rho_s,z_s})=\int_{0}^{\theta_0}d\theta\,\sin\theta   (\cos\theta)^{3/2}\,e^{-\gamma_f^2\sin^2\theta}d_{m,\sigma}^{\ell}(\theta)\, J_{m-\sigma}\left( k \rho_s \sin\theta\right) e^{i\,k\cos\theta\, z_s }, 
\label{C5multipole coefficient}
\end{equation}

\begin{equation}
G_{\ell,m}^{(\sigma)\pm}({ \rho_s,z_s})=\int_{0}^{\theta_0}{ d\theta}\, \sin^2\theta \sqrt{\cos \theta}\,e^{-\gamma_f^2\sin^2\theta}d_{m\pm 1,\sigma}^{\ell}(\theta)\, J_{m-\sigma}\left( k \rho_s \sin\theta\right) e^{i\,k\cos\theta\, z_s } \, ,
\end{equation}

\end{widetext}
 where 
 $k=n_wk_0$ is the
 wavenumber in 
 the aqueous host medium. 
 The corresponding wavevector makes an angle $\theta$
 with respect to the optical $z$-axis.

\end{document}